\documentclass[aps,prd,nofootinbib,amsmath,amssymb,superscriptaddress,twocolumn,10pt]{revtex4}

\pdfoutput=1

\usepackage{graphicx}
\usepackage{dcolumn}
\usepackage{bm}
\usepackage{amssymb}
\usepackage{latexsym}
\usepackage{booktabs}
\usepackage{amsmath}
\usepackage{multirow}
\usepackage[colorlinks=true, linkcolor=red, citecolor=blue]{hyperref}

\newcommand{\be}{\begin{equation}}
\newcommand{\ee}{\end{equation}}
\newcommand{\bq}{\begin{eqnarray}}
\newcommand{\eq}{\end{eqnarray}}

\usepackage[usenames,dvipsnames]{xcolor}

\begin{document}

\title{Improving cosmological parameter estimation with the future gravitational-wave standard siren observation from the Einstein Telescope}

\author{Xuan-Neng Zhang}
\affiliation{Department of Physics, College of Sciences, Northeastern
University, Shenyang 110819, China}
\author{Ling-Feng Wang}
\affiliation{Department of Physics, College of Sciences, Northeastern
University, Shenyang 110819, China}
\author{Jing-Fei Zhang}
\affiliation{Department of Physics, College of Sciences, Northeastern
University, Shenyang 110819, China}
\author{Xin Zhang\footnote{Corresponding author}}
\email{zhangxin@mail.neu.edu.cn}
\affiliation{Department of Physics, College of Sciences, Northeastern
University, Shenyang 110819, China} 
\affiliation{Ministry of Education Key Laboratory of Data Analytics and Optimization
for Smart Industry, Northeastern University, Shenyang 110819, China}
\affiliation{Center for High Energy Physics, Peking University, Beijing 100080, China}
\affiliation{Center for Gravitation and Cosmology, Yangzhou University, Yangzhou 225009, China}

\begin{abstract}

Detection of gravitational waves produced by merger of binary compact
objects could provide an independent way for measuring the luminosity
distance to the gravitational-wave burst source, indicating that
gravitational-wave observation, combined with observation of
electromagnetic counterparts, can provide ``standard sirens'' for
investigating the expansion history of the universe in cosmology. In this
work, we wish to investigate how the future gravitational-wave standard
siren observations would break the parameter degeneracies existing in the
conventional optical observations and how they help improve the parameter
estimation in cosmology. We take the third-generation ground-based
gravitational-wave detector, the Einstein Telescope, as an example to make
an analysis. By simulating 1000 events data in the redshift range between
0 and 5 based on the ten-year observation of the Einstein Telescope, we
find that the gravitational-wave data could largely break the degeneracy
between the matter density and the Hubble constant, thus significantly
improving the cosmological constraints. We further show that the
constraint on the equation-of-state parameter of dark energy could also be
significantly improved by including the gravitational-wave data in the
cosmological fit.

\end{abstract}
\maketitle

The observations from the Planck satellite mission strongly favor a
6-parameter base $\Lambda$ cold dark matter ($\Lambda$CDM) cosmology
\cite{Aghanim:2015xee}. That is to say, in the current stage, one can use
only 6 parameters to reproduce the evolution of the universe, including the
expansion history and the large-scale structure formation, based on the
$\Lambda$CDM model, which is in good agreement with the current various
cosmological observations. However, it is hard to believe that the eventual
model of cosmology indeed consists of only 6 parameters. Actually, one
believes that the current status is due to the fact that the current observations are not precise
enough to tightly constrain other possible parameters beyond the base
$\Lambda$CDM model, and in the future the base model must be extended in
several aspects with the help of future highly accurate observational data.

In fact, it seems that some cracks appear in the base $\Lambda$CDM
cosmology, which is hinted by the fact that some tensions exist between
different astrophysical observations based on the base $\Lambda$CDM model,
e.g., the well-known issues concerning $H_0$ and $\sigma_8\Omega_{\rm
m}^\alpha$ (with $\alpha$ taken to be 0.3--0.5) measurements
\cite{Aghanim:2015xee, Ade:2013zuv,Mortonson:2013zfa}. A possible way to
solve the tensions is to consider some extensions to the base $\Lambda$CDM
model, but this leads to the fact that the extra parameters are rather
difficult to be tightly constrained by the current observational data, in
particular, the parameter degeneracies might be strong for these extra
parameters. All these facts actually are making requests for the current
cosmology: (i) cosmological probes should be further developed; and (ii)
cosmological model should also be further extended.

Currently, the major cosmological probes mainly include: cosmic microwave
background anisotropies (CMB), type Ia supernovae (SN), baryon acoustic oscillations (BAO),
direct determination of the Hubble constant, weak gravitational lensing,
clusters of galaxies, and redshift-space distortions. The combinations of
these cosmological data have provided precise measurements for some
cosmological parameters, but for several important extra parameters, e.g.,
the equation-of-state parameter of dark energy $w(z)$ (with $z$ being
redshift), the total mass of neutrinos $\sum m_\nu$, the effective number of
relativistic species $N_{\rm eff}$, and so forth, the current observations
still cannot provide tight constraints \cite{Aghanim:2015xee}. The future
major dark energy experiments (e.g., DESI \cite{desi}, LSST \cite{lsst}, Euclid \cite{euclid}, WFIRST \cite{wfirst}, etc.) will
definitely play a crucial role in determining these parameters, but all
these experiments are optical (or near-infrared imaging or spectroscopy)
observations, which implies that any new observational means would be
helpful in avoiding the systematic errors in these optical observations. The
promising new cosmological probes mainly include the radio observations
(i.e., 21 cm neutral hydrogen survey) and the gravitational-wave
observations.

It is well known from Schutz's work in the mid-1980s \cite{Schutz1986} that the gravitational
waves carry information of cosmic distance of the source.
Actually, from the observation of the waveform of gravitational waves
released by binary compact objects merger, one can independently measure the
luminosity distance to the source of gravitational-wave burst. Furthermore,
if the redshift of the source can also be observed by identifying the
electromagnetic (EM) counterpart of the source (like the cases of the merger
of two neutron stars and the merger of a black hole and a neutron star), then
one can establish a true distance-redshift relation based on the
observation of large amount gravitational-wave events from which the
expansion history of the universe can be inferred
\cite{Holz:2005df,Dalal:2006qt,MacLeod:2007jd,Sathyaprakash:2009xt,Taylor:2011fs,DelPozzo:2011yh,DelPozzo:2015bna,Cai:2017cbj,prx}.
Compared with the observation of type Ia SN that can also
measure the luminosity distance in some sense, the gravitational-wave (GW)
observation as a cosmological probe has the following advantages: (i) The SN
observation actually cannot measure the absolute luminosity distance, but
can only measure the ratio of luminosity distances at different redshifts,
due to the fact that the intrinsic luminosity of type Ia SN is not precisely
known to us. But the GW observation, definitely, can provide measurement for
the absolute luminosity distance to the source. Note also that the measurement of luminosity distance by GW observation is independent of a cosmic distance ladder. (ii) The SN observation can
only provide measurements for events with redshifts less than about 1.4, but the
GW observation can provide measurements for events with much higher
redshifts. These advantages ensure that the future GW observations could
play a significant role in breaking the parameter degeneracies and help
improve the cosmological parameter estimation.

It should be mentioned here that the detection of the event GW170817 \cite{TheLIGOScientific:2017qsa}, a strong signal from the merger of a binary neutron-star system, by the Advanced LIGO and Virgo detectors demonstrated that the era of multi-messenger astronomy has begun. The identification of NGC 4993 as the host galaxy of GW170817 enables us to perform a standard siren measurement of the Hubble constant. Such an independent measurement of the Hubble constant gives the result of $H_0=70.0^{+12.0}_{-8.0}$ km s$^{-1}$ Mpc$^{-1}$ \cite{Abbott:2017xzu}, which is broadly consistent with the existing measurements. This first GW-EM multi-messenger event demonstrates that GW standard siren observations have potential for cosmological inference. 

Some forecast studies on cosmology by using the future GW observations (with
short-hard $\gamma$-ray bursts or other EM counterparts, as standard sirens) have
been performed in the literature (see, e.g.,
Refs.~\cite{Nissanke:2009kt,Zhao:2010sz,Tamanini:2016zlh,Cai:2016sby,Cai:2017yww,Cai:2017plb,Zhao:2017cbb,Cai:2017aea,Feeney:2018mkj,Wang:2018lun}).
For example, in Ref.~\cite{Nissanke:2009kt}, it is shown that the
observation from a network of advanced LIGO detectors can constrain the
Hubble constant to a 5\% accuracy. In Ref.~\cite{Zhao:2010sz}, it is
demonstrated that the observation of 1000 GW events from the next-generation
ground-based GW detector, the Einstein Telescope (ET), is possible to
constrain the cosmological parameters up to $\sigma(h)\sim 5\times 10^{-3}$
and $\sigma(\Omega_{\rm m})\sim 0.02$ using a Fisher information matrix approach; see
also Ref.~\cite{Cai:2016sby} in which similar results are found using a Markov-chain Monte Carlo (MCMC)
method. In Ref.~\cite{Cai:2016sby}, it is found that using the Gaussian
Process method the equation-of-state parameter of dark energy can be
constrained to be $\sigma[w(z)]\sim 0.03$ in the low redshift region with the future GW observations.
Also, in Ref.~\cite{Wang:2018lun}, it is shown that with the help of 1000 GW
events observed from the ET the constraints on the neutrino mass can be
improved by about 10\%. Furthermore, based on the space-based detector LISA,
the expansion of the universe and the interacting dark energy have also been
investigated in Refs.~\cite{Cai:2017yww,Cai:2017plb}.

In this work, we wish to investigate how the future GW standard siren
observation would break the cosmological parameter degeneracies (existing in
the conventional observations) and thus help improve the parameter
estimation for cosmology. We will take the third-generation ground-based
detector ET as an example to make an analysis. In this analysis, to be in
accordance with the previous studies ~\cite{Zhao:2010sz, Cai:2016sby,
Wang:2018lun}, we will simulate 1000 GW events data in the redshift range of
$z\in [0,5]$ based on the ET's ten-year observation.\footnote{Here we briefly explain why we 
consider to simulate 1000 GW events data (with EM counterparts). According to the Conceptual Design Study of the Einstein Telescope \cite{ET} (see Table 2 on Page 31), the event rate (yr$^{-1}$) in ET is ${\cal O}(10^3-10^7)$ for coalescences of binary neutron stars (BNS) [and also for neutron star-black hole (NS-BH)]. Following Ref.~\cite{Zhao:2010sz}, we here take the event rate in ET to be $10^5$ per year for BNS and NS-BH. Considering that short gamma-ray bursts (SGRBs) generated by coalescences of BNS or NS-BH are believed to be beamed with small beaming angle, only a small fraction of the total events are expected to be observed as SGRBs. The fraction for the ``useful'' events (i.e., the GW events accompanied with the observation of SGRB) is assumed to be $10^{-3}$ \cite{Cai:2016sby}. Thus, the number of simulated GW standard siren data considered in this work is $10^5~{\rm yr}^{-1}\times 10^{-3}\times 10~{\rm yr}=10^3$. See also, e.g., Refs.~\cite{Zhao:2010sz, Cai:2016sby,Wang:2018lun}.}
The GW data simulation
method used in this paper is in exact accordance with the prescription given
in Refs.~\cite{Cai:2016sby, Wang:2018lun}, and thus we do not repeat it
here; we refer the reader to Refs.~\cite{Cai:2016sby, Wang:2018lun} for
details.\footnote{Note here that in the simulation we mainly consider the coalescence events of BNS and also consider a small number of NS-BH coalescence events. According to the prediction of the Advanced LIGO-Virgo network, we take the ratio of the events numbers of NS-BH and BNS to be 0.03. For the mass distributions of NS and BH, we randomly sample the mass of NS in the interval of $[1,2]~M_\odot$ and the mass of BH in the interval of $[3,10]~M_\odot$, where $M_\odot$ is the solar mass, in accordance with Refs.~\cite{Cai:2016sby, Wang:2018lun}.} 
There are some parameter degeneracies in cosmological models
constrained by the current conventional observations, such as CMB, BAO, and SN. Using the GW data to make parameter estimation, there will
also be some degeneracies, but in this case the orientation of the
degeneracy in some parameter plane would be different from the above case.
Thus, the GW standard siren observation will play an important role in
improving the parameter estimation because it could break
parameter degeneracies in the conventional observations. In this paper, we
will take the $\Lambda$CDM model and the $w$CDM model [where the
equation-of-state parameter (EoS) of dark energy $w$ is taken to be a constant] as
examples to see how this happens.

\begin{figure*}[!htp]
\includegraphics[scale=0.3]{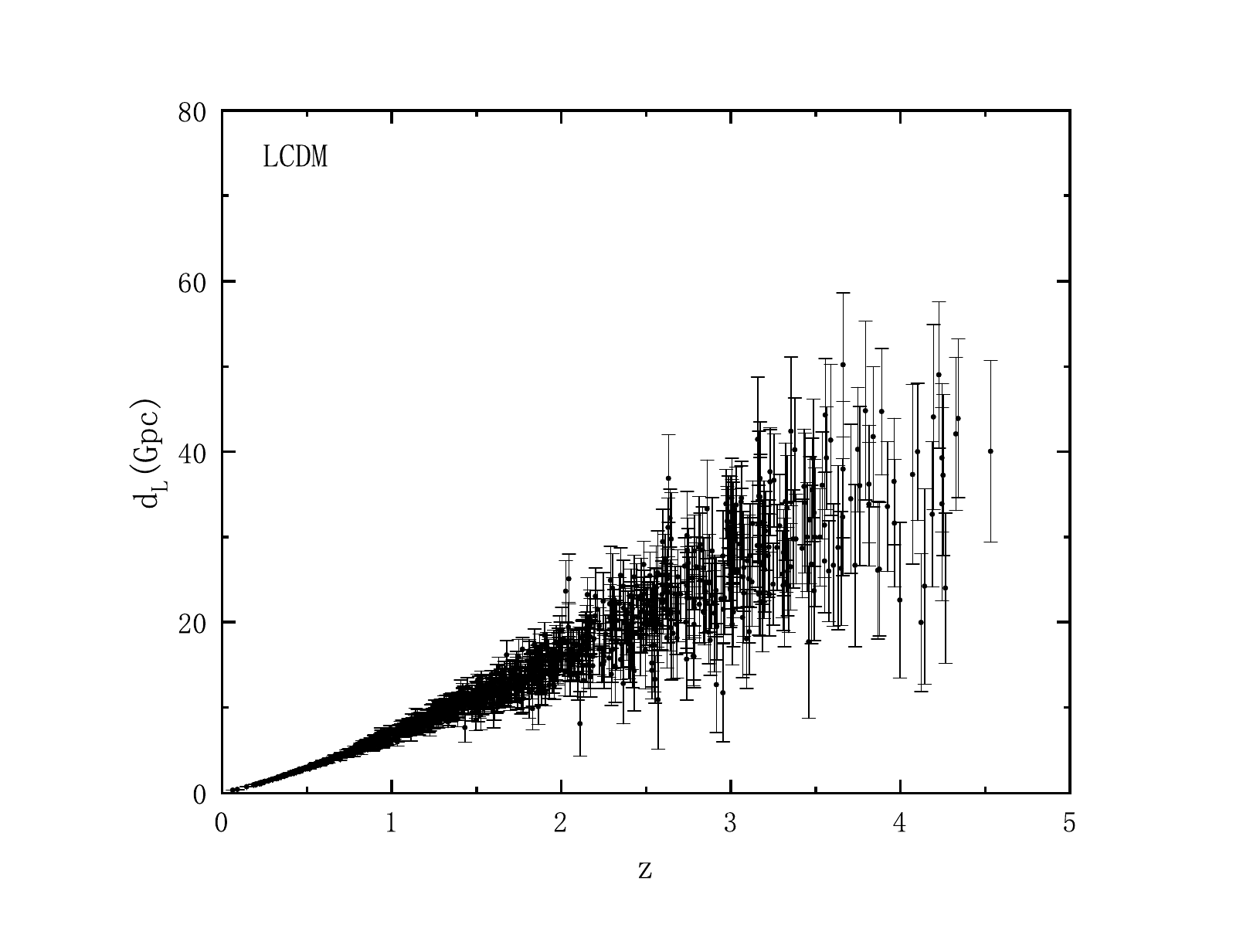}
\includegraphics[scale=0.3]{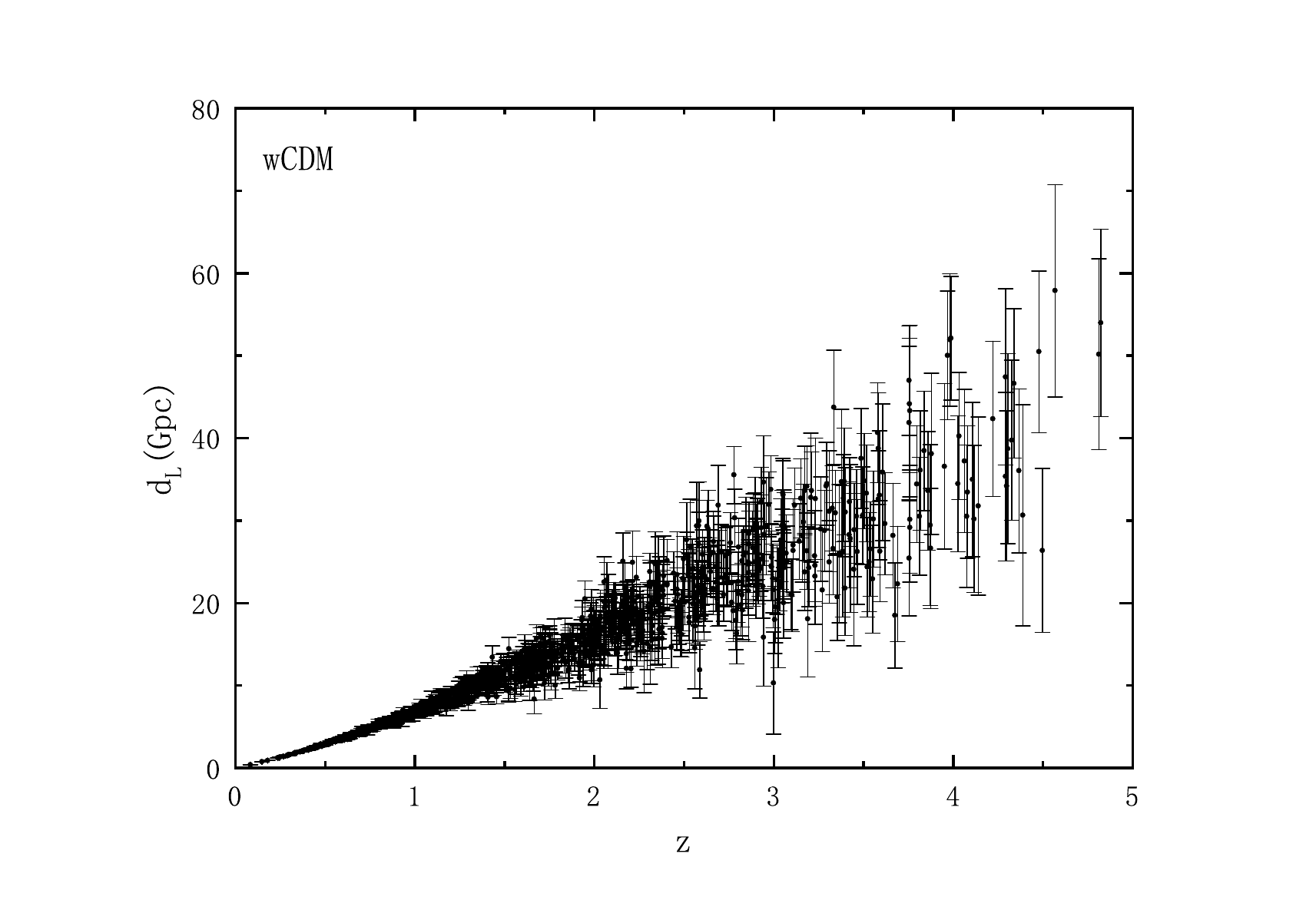}
\centering
\caption{\label{fig1} The luminosity distance data points from simulated 1000 gravitational-wave events in the redshift range of $z\in [0,5]$ based on the ten-year observation of the Einstein Telescope. The fiducial models are chosen as the best-fitted $\Lambda$CDM ({\it left}) and $w$CDM ({\it right}) models, respectively, constrained by the current CMB+BAO+SN data.}
\end{figure*}

\begin{table*}[!htp]
        \small
        \centering
        \caption{\label{tab1} Fitting results for the $\Lambda$CDM model and the $w$CDM model using CBS, GW, and CBS+GW. Here, CBS stands for CMB+BAO+SN.}

		\begin{tabular}{cccccccccc}
\hline
\hline Model&&\multicolumn{3}{c}{$\Lambda$CDM}&&\multicolumn{3}{c}{$w$CDM}&\\
        \cline{1-1}\cline{3-5}\cline{7-9}
 Data&&CBS&GW&CBS+GW &&CBS&GW&CBS+GW&\\
       \hline

 $\Omega_{\rm b}h^2$&&$0.02233\pm0.00014$&$0.050^{+0.050}_{-0.045}$&$0.02229\pm0.00010$&&$0.02230\pm0.00015$&$0.051^{+0.049}_{-0.046}$&$0.02231\pm0.00013$\\
 $\Omega_{\rm c}h^2$&&$0.11853\pm0.00101$&$0.091^{+0.038}_{-0.025}$&$0.11898^{+0.00050}_{-0.00051}$&&$0.11893^{+0.00124}_{-0.00125}$&$0.081^{+0.036}_{-0.029}$&$0.11872^{+0.00105}_{-0.00104}$\\
 $100\theta_{\rm{MC}}$&&$1.04092\pm0.00030$&$0.96^{+0.12}_{-0.11}$&$1.04086^{+0.00027}_{-0.00026}$&&$1.04087\pm0.00030$&$0.95\pm0.11$&$1.04088\pm0.00028$\\
 $\tau$&&$0.086\pm0.016$&--&$0.083\pm0.016$&&$0 .084\pm0.017$&--&$0 .084\pm0.017$\\
 $n_{\rm s}$&&$0.9679^{+0.0039}_{-0.0040}$&--&$0.9666^{+0.0032}_{-0.0033}$&&$0 .9668\pm0.0044$&--&$0 .9673\pm0.0040$\\
 ${\rm{ln}}(10^{10}A_{\rm s})$&&$3.103\pm0.032$&--&$3.098\pm0.032$&&$3.101\pm0.033$&--&$3 .101\pm0.033$\\
 $\Omega_{\rm m}$&&$0.3075^{+0.0060}_{-0.0061}$&$0.3119^{+0.0071}_{-0.0073}$&$0.3103\pm0.0024$&&$0.3039^{+0.0094}_{-0.0093}$&$0.2890\pm0.0150$&$0.3061\pm0.0023$\\
 $H_0\,[{\rm km}/{\rm s}/{\rm Mpc}]$&&$67.84\pm0.46$&$67.57\pm0.22$&$67.62\pm0.16$&&$68.3^{+1.0}_{-1.1}$&$67.79^{+0.32}_{-0.33}$&$68.04^{+0.23}_{-0.24}$\\
 $w$&&--&--&--&&$-1.023^{+0.042}_{-0.041}$&$-0.943^{+0.062}_{-0.054}$&$-1.010\pm0.020$\\
\hline \hline
		\end{tabular}
		
\end{table*}

\begin{table}[!htp]
        \small
        \centering
\caption{\label{tab2} Constraint errors and accuracies for parameters of $\Lambda$CDM and $w$CDM using CBS, GW, and CBS+GW. Here, CBS stands for CMB+BAO+SN.}
		\begin{tabular}{ccccccccc}
\hline
\hline Model&&\multicolumn{3}{c}{$\Lambda$CDM}&&\multicolumn{3}{c}{$w$CDM}\\
\cline{1-1}\cline{3-5}\cline{7-9}
 Data&&CBS&GW&CBS+GW&&CBS&GW&CBS+GW\\
\hline
 $\sigma(\Omega_{\rm m})$&&0.0060&0.0072&0.0024&&0.0094&0.0150&0.0023\\
 $\sigma(h)$&&0.0046&0.0022&0.0016&&0.0105&0.0033&0.0024\\
 $\sigma(w)$ &&--&--&--&&0.042&0.058&0.020\\
 $\varepsilon(\Omega_{\rm m})$&&0.0197&0.0231&0.0077&&0.0308&0.0520&0.0075\\
 $\varepsilon(h)$&&0.0068&0.0033&0.0024&&0.0153&0.0048&0.0035\\
 $\varepsilon(w)$&&--&--&--&&0.041&0.062&0.020\\
\hline
\hline
\end{tabular}

\end{table}
\begin{figure*}[!htp]
\includegraphics[scale=0.9]{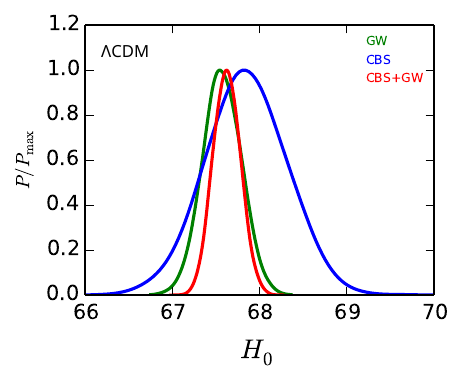}
\includegraphics[scale=0.9]{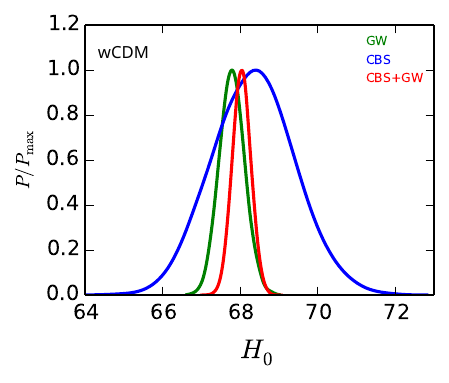}
\centering
\caption{\label{fig2} One-dimensional posterior distributions of $H_0$ in the $\Lambda$CDM model ({\it left}) and the $w$CDM model ({\it right}) using the CBS, GW, and CBS+GW data combinations. Here, CBS stands for CMB+BAO+SN.}
\end{figure*}

\begin{figure*}[!htp]
\includegraphics[scale=0.9]{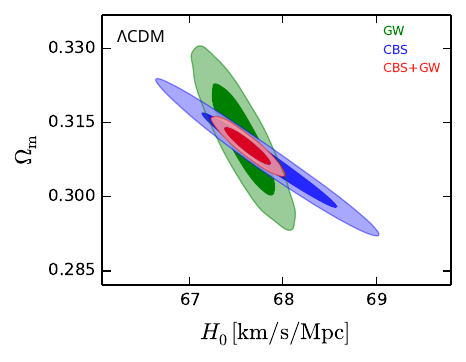}
\includegraphics[scale=0.9]{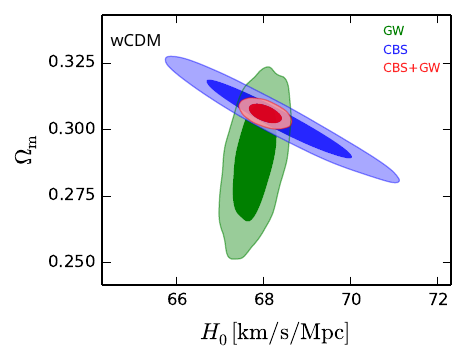}
\centering
 \caption{\label{fig3} Constraints (68.3\% and 95.4\% CL) on the $\Lambda$CDM model ({\it left}) and the $w$CDM model ({\it right}) in the $H_0$--$\Omega_{\rm m}$ plane using the CBS, GW, and CBS+GW data combinations. Here, CBS stands for CMB+BAO+SN.}
\end{figure*}
\begin{figure}[!htp]
\includegraphics[scale=0.9]{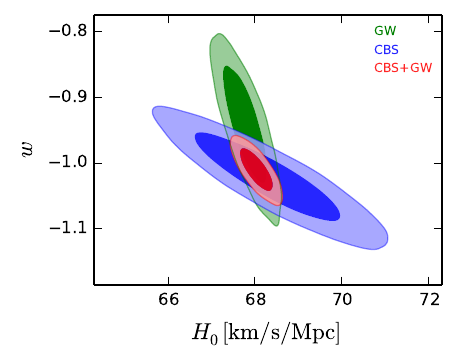}
\centering
 \caption{\label{fig4} Constraints (68.3\% and 95.4\% CL) on the $w$CDM model in the $H_0$--$w$ plane using the CBS, GW, and CBS+GW data combinations. Here, CBS stands for CMB+BAO+SN.}
\end{figure}

First, we will use the current observations of CMB, BAO, and SN to constrain the two cosmological models, and see how the parameters degenerate with each other, leading to the result that they cannot be well constrained. Then, we will use the simulated GW data from the ET to constrain the models, and we will observe different degeneracy cases, which leads to the fact that the previous degeneracies are broken when the GW data are included in the data combination.

We now briefly describe the current observations used in this paper. For CMB
data, we use the Planck temperature and polarization power spectra (Planck
TT,TE,EE+lowP) \cite{Aghanim:2015xee}. For BAO data, we use the measurements
from 6dFGS ($z_{\rm eff}=0.106$) \cite{Beutler:2011hx}, SDSS-MGS ($z_{\rm
eff}=0.15$) \cite{Ross:2014qpa}, and BOSS LOWZ ($z_{\rm eff}=0.32$) and
CMASS ($z_{\rm eff}=0.57$) \cite{Cuesta:2015mqa}. For SN data, we use the
JLA compilation \cite{Betoule:2014frx}. We use the data combination of
CMB+BAO+SN to constrain the $\Lambda$CDM and $w$CDM models by employing the
MCMC package {\tt CosmoMC} \cite{Lewis:2002ah}, and
then take the best-fitted models as the fiducial models to generate the
simulated GW data. In Fig.~\ref{fig1}, we show the simulated GW data
consisting of $d_L^{\rm mea}$ and $\sigma_{d_L}$ in the redshift range of
$z\in [0,5]$ for the two cases in which the $\Lambda$CDM model and the
$w$CDM model are taken as fiducial models, respectively.

We then use the simulated GW data to constrain the models. The constraints from the data combination of CMB+BAO+SN+GW will show how the GW data help improve the parameter estimation in the considered two cases. We summarize the fitting results in Tables~\ref{tab1} and \ref{tab2}. In Table~\ref{tab1} we show the fitting values of cosmological parameters, and in Table~\ref{tab2} we show the constraint errors and constraint accuracies for the concerned parameters (i.e., $\Omega_{\rm m}$, $H_0$, and $w$). Here, the error $\sigma$ is taken to be the average of $\sigma_+$ and $\sigma_-$, and $\varepsilon(\xi)$ for a parameter $\xi$ is defined as $\varepsilon(\xi)=\sigma(\xi)/\xi$.

In Fig.~\ref{fig2}, we show the one-dimensional posterior distributions of $H_0$ in the $\Lambda$CDM model ({\it left}) and the $w$CDM model ({\it right}) using the CMB+BAO+SN, GW, and CMB+BAO+SN+GW data combinations. Note that, for convenience, hereafter we use the abbreviation ``CBS'' to denote the combination CMB+BAO+SN. Here we can clearly see that the GW observation solely can tightly constrain the Hubble constant. In the case of $\Lambda$CDM, we have $\sigma(h)=4.6\times 10^{-3}$ and $\varepsilon(h)=0.68\%$ from CBS, $\sigma(h)=2.2\times 10^{-3}$ and $\varepsilon(h)=0.33\%$ from GW, and $\sigma(h)=1.6\times 10^{-3}$ and $\varepsilon(h)=0.24\%$ from CBS+GW. We find that the accuracy of $H_0$ is improved from 0.68\% to 0.24\% in the $\Lambda$CDM case when the GW data are included in the fit. In the case of $w$CDM, we have $\sigma(h)=1.05\times 10^{-2}$ and $\varepsilon(h)=1.53\%$ from CBS, $\sigma(h)=3.3\times 10^{-3}$ and $\varepsilon(h)=0.48\%$ from GW, and $\sigma(h)=2.4\times 10^{-3}$ and $\varepsilon(h)=0.35\%$ from CBS+GW. We find that the accuracy of $H_0$ is improved from 1.53\% to 0.35\% in the $w$CDM case when the GW data are included in the fit. Obviously, it is shown from this analysis that the GW observation is capable of significantly improve the constraint accuracy of $H_0$, in particular in dynamical dark energy models.

Figure~\ref{fig3} shows constraints on the $\Lambda$CDM model ({\it left}) and the $w$CDM model ({\it right}) in the $H_0$--$\Omega_{\rm m}$ plane. In this figure, we can clearly see that, from the CBS constraint, in both the $\Lambda$CDM and $w$CDM models, $\Omega_{\rm m}$ and $H_0$ are in strong anti-correlation. In the case of $\Lambda$CDM, the GW constraint still gives an anti-correlation for $\Omega_{\rm m}$ and $H_0$, but its degeneracy orientation in the parameter plane is evidently different from the former, resulting in the breaking of the degeneracy. In the case of $w$CDM, we find that the GW constraint leads to a weak positive correlation for $\Omega_{\rm m}$ and $H_0$, and thus the orthogonality of the two degeneracy orientations results in a complete breaking of the parameter degeneracy. Thus, although CBS and GW give similar constraints on $\Omega_{\rm m}$, the combination of the two could tremendously improve the constraint on $\Omega_{\rm m}$. In the $\Lambda$CDM case the constraint on $\Omega_{\rm m}$ is improved from 1.97\% to 0.77\%, and in the $w$CDM case the constraint on $\Omega_{\rm m}$ is from 3.08\% to 0.75\%, when the GW data are included in the fit.

In Fig.~\ref{fig4}, we show the constraints on the $w$CDM model in the $H_0$--$w$ plane. From this figure, we can also clearly see that the parameter degeneracy could be largely broken by including the GW observation in the cosmological fit. The CBS data give $\sigma(w)=0.042$ and the GW data give $\sigma(w)=0.058$, indicating that their constraints on $w$ are similar. But the combination of the two gives $\sigma(w)=0.020$, showing that the error is largely decreased. By considering the GW observation, the constraint accuracy of $w$ is improved from 4.1\% to 2.0\%. Therefore, in this analysis, we have shown that the GW observation could help greatly improve the constraint accuracy of $w$.

Finally, let us make some relevant discussions.
GW standard sirens have some advantages in studying cosmology. For example, GW standard sirens are self-calibrating and thus have no dependence on a cosmic distance ladder, which leads to that observation of GW standard sirens can measure true luminosity distances and establish a true distance-redshift relation to study cosmography. Moreover, compared with SN observation, GW observation can measure sources with higher redshifts. Although we know that GW standard sirens have these advantages, we still need to confirm the usefulness of GW standard sirens in precision cosmology. The core aim of this work is to investigate what role the future observation of GW standard sirens can play in the study of precision cosmology. 

Compared with previous studies in the literature, the feature of this work is to be with a view to the question of whether the GW observation from ET can break the parameter degeneracies existing in the EM observations. In Ref.~\cite{Zhao:2010sz} it was found by using a Fisher matrix approach that ET (combined with the Planck CMB prior) will be able to constrain $w_0$ and $w_a$ [for the dynamical dark energy model with parameterization $w(a)=w_0+w_a(1-a)$ with $a$ being the scale factor of the universe] with the errors $\sigma(w_0)=0.099$ and $\sigma(w_a)=0.302$, respectively, and these results are comparable with the projected errors for the JDEM BAO project and the SNAP SN observations. In Ref.~\cite{Cai:2016sby} it was found by using a MCMC approach that the Gaussian Process method can help reconstruct the evolution of EoS of dark energy and the future GW observation from ET would give $\sigma[w(z)]\sim 0.03$ in the low redshift region. However, in the present study we wish to focus on the issue of the role of GW observation in precision cosmology. We investigate what the role the future ET GW observation would play in constraining cosmological parameters in combination with EM observations. We only employ the simplest cosmological models, i.e., the $\Lambda$CDM model and the $w$CDM model, to complete the analysis, in which a MCMC method is used. It is shown in this work that some parameter degeneracies can be successfully broken by using the GW observation. Although the GW observation can be used to precisely determine the Hubble constant $H_0$, for other cosmological parameters the GW observation actually cannot provide very tight constraints. For example, the GW observation can only constrain $\Omega_{\rm m}$ and $w$ (in the $w$CDM model) at the accuracies of 5.20\% and 6.2\%, respectively, while the current CMB+BAO+SN observation can constrain them at the accuracies of 3.08\% and 4.1\%, respectively, actually better than the former. However, owing to the fact that the degeneracy orientations from GW and EM observations in the parameter space are distinctively different from each other, the parameter degeneracies can be well broken, which leads to that the combined data of GW and EM can constrain $\Omega_{\rm m}$ and $w$ at the accuracies of 0.75\% and 2.0\%, respectively. Therefore, the goal of this work is achieved: We have concluded from our analysis that the role of the future ET GW observation in the precision cosmology is to break the parameter degeneracies of the EM observations, and the combination of GW and EM observations can provide much better constraints on cosmological parameters.

Actually, using only a small number of low-redshift GW events produced by BNSs can precisely determine the value of the Hubble constant $H_0$ in a model-independent way. This will definitely resolve the $H_0$ tension and judge if the $\Lambda$CDM model should be extended. A recent study \cite{Feeney:2018mkj} pointed out that about 50 BNS standard sirens data will be able to fulfill this task. In the present study we have pointed out that in extensions to $\Lambda$CDM cosmology the parameter degeneracies (in particular for the extra parameters) can be well broken by the future GW standard sirens observation. It is expected that in the future the combination of GW and EM observations would push the precision cosmology forward by a great step.

In summary, it is shown in this work that the future GW standard siren observation is capable of breaking the parameter degeneracies existing in the conventional optical observations and thus could help improve the parameter estimation for cosmology. We have simulated 1000 GW events data based on the ET's ten-year observation. In order to show how the GW data break the parameter degeneracies, we employ the current CMB+BAO+SN data to make comparison and combination. We take the $\Lambda$CDM and $w$CDM models as examples to make an analysis. We find that the degeneracy between $\Omega_{\rm m}$ and $H_0$ can be greatly broken by including the GW data, particularly for the case of $w$CDM. Thus, for both $\Omega_{\rm m}$ and $H_0$, the constraint accuracies are tremendously improved by considering the GW data from ET. Although $w$ is hard to be tightly constrained, the GW observation from ET could help improve its constraint accuracy from 4\% to 2\%, according to our analysis. In this work, we only make a preliminary analysis, because we only consider the improvement for the parameter estimation based on the current CMB+BAO+SN data, but not the future optical observations. Moreover, we only consider the simplest dynamical dark energy model, i.e., the $w$CDM model, in this work. We will leave a comprehensive analysis to a future work.

\begin{acknowledgments}
We would like to thank Zhou-Jian Cao, Tao Yang, and Wen Zhao for helpful discussions. This work was supported by the National Natural Science Foundation of China (Grants Nos.~11875102,
11835009, 11522540, and 11690021) and the National Program for Support of Top-Notch Young Professionals.


\end{acknowledgments}



\end{document}